# Radiative recombination through EL2 centers in gallium arsenide single crystals doped by selenium and cadmium

*M.B. Litvinova, S.V. Shutov\*, A.D. Shtan'ko\*\*, V.V. Kurak\*\**

National University of Shipbuilding,
44 Ushakova Ave., 73022, Kherson, Ukraine
\*V.Lashkaryov Institute of Semiconductor Physics,
National Academy of Sciences of Ukraine,
41 Nauki Ave., 03028, Kyiv, Ukraine
\*\*Kherson National Technical University,
24 Berislavskoe Road, 73008, Kherson, Ukraine

Influence of Cd and Se atoms on the quantum efficiency of photon emission through EL2 defects in gallium arsenide single crystals has been investigated. A comparative technique of impurity diffusion in vacuum and arsenic atmospheres has been used. The change character and extent of the photon emission quantum efficiency have been established to be defined by vacancy structure of crystal that is most likely caused by formation of EL2-dopant complexes.

Optical methods of deep centers investigation in GaAs provide information on the recombination process nature as well as on the structure of crystal defects. In this work, a number of factors is established causing the quantum efficiency changes of radiative transitions through anti-structural ($As_{Ga}$) EL2 defects under doping of semi-insulating undoped (SIU) GaAs single crystals.

The recombination processes involving the EL2 centers are known to induce appearing of an emission band with energy maximum at $h\nu_m \approx 0.65$ eV in low temperature photoluminescence (PL) spectra. This band is a superposition of bands at $h\nu_m \approx 0.63$ eV and $h\nu_m \approx 0.68$ eV (Fig. 1) [1]. The first band is due to radiative capture of free electrons by the charged $EL2^+$ defects while the second one, to radiative capture of free holes by the neutral $EL2°$ defects (see inset to Fig. 1, the transitions 2 and 3 respectively). The efficiency of recombination processes through $EL2^+$ and $EL2°$ centers is influenced by a number of factors, first of all, by the self-diffusion processes. Diffusion of arsenic vacancies from crystal surface during vacuum high-temperature annealing (TA) of the material causes decreasing of EL2 center concentration ($N_{EL2}$) due to the reaction

$$As_{Ga} + V_{As} \Leftrightarrow V_{Ga} + As_{As} \qquad (1)$$

and decreasing intensity of the $h\nu_m = 0.65$ eV ($I_{0.65}$) radiation. On the contrary, formation of new EL2 centers during TA under excess As vapor pressure ($p_{As} > 9.8 \cdot 10^4$ Pa) results in an increased $I_{0.65}$ [2]. A sharp drop of $I_{0.65}$ is also observed in copper doping of the GaAs crystals [3].

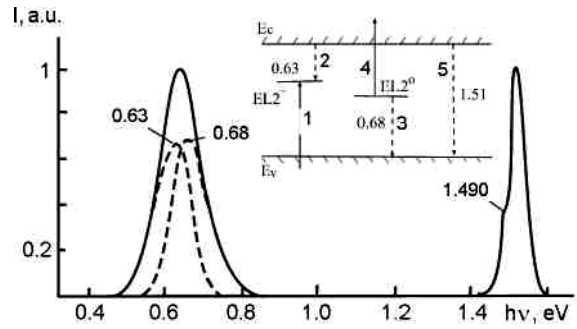

Fig. 1. PL spectra of SIU GaAs crystals at 77 K. Dotted lines show expansion of $h\nu_m \approx 0.65$ eV band into components with maxima at 0.63 and 0.68 eV. Inset: energy scheme of corresponding optical transitions (solid lines correspond to absorption, dotted ones, to emission of light quantum).

This effect is caused by the passivation of the EL2 defects resulting from formation of electrically inactive EL2 – Cu complexes. Besides, copper introduction causes a decreased luminescence quantum yield through neutral $EL2^0$ centers (as compared to PL quantum yield for luminescence through the charged $EL2^+$ centers). In [4, 5], it has been shown that changes in the intensity ratio of $h\nu_m = 0.63$ eV to $h\nu_m = 0.68$ eV bands is influenced by the vacancy composition of the crystal and that in EL2—Cu complexes, copper atoms occupy mainly Ga vacancies (EL2 – $Cu_{Ga}$).

It can be supposed that the decreased $I_{0.65}$ intensity in SIU GaAs crystals as a result of EL2-dopant complex formation can be supposed to be due not only to copper atoms but also to atoms of other impurities. Furthermore, there is a question about influence of the positional state of atoms in the crystal (i.e. either the dopant atoms have occupied positions in gallium or arsenic sublattice) on this process. In this work, the changes in radiative recombination through EL2 centers under doping of GaAs crystals by Cd atoms which diffuse involving arsenic vacancies and by selenium atoms diffused mainly through gallium vacancies [6] is studied.

The n-type SIU GaAs (100) crystals grown by the Czochralsky technique with resistivity $p = 7 \cdot 10^7 - 2 \cdot 10^8$ $\Omega \cdot$cm were used as initial material in our experiments. The EL2 center concentration, as determined from optical absorption of $h\nu = 1.04$ eV light quanta [1] was $N = N^0 + N^+ = (1.2$ to $2.0) \cdot 10^{16}$ cm$^3$. The vacancy composition of crystals (concentration ratio between arsenic and gallium vacancies, $z = [V_{As}]/[V_{Ga}]$ was determined from intensity ratio between PL bands at 77 K: the edge band with the emission maximum at $h\nu_m = 1.51$ eV and extrinsic one, with $h\nu_m = 1.49$ eV, caused by radiative transition from the conductivity band to the acceptor level $C_{As}$ (Fig. 1) [7]. As a radiation source for PL excitation, a He-Ne laser with radiation intensity of $(2.5-3.0) \cdot 10^{18}$ cm$^{-2}$s$^{-1}$ was used. The PL spectra were recorded by a standard technique [8].

Selenium and cadmium diffusion was carried out by annealing of the GaAs samples at 800°C for 6 h in quartz ampoules in vacuum as well as under excess pressure of arsenic vapor. As the source of diffusing atoms, a 1 μm thick Se or Cd layer thermally sputtered onto the sample surface was used. To realize the diffusion process, the samples of about 2 mm thickness were placed in a quartz ampoule of about 4 cm$^3$ operating volume. The ampoules, pre-degreased and treated in aqua regia to reduce uncontrollable contamination of crystals with copper, were evacuated to a residual pressure less than $10^{-3}$ Pa. A 20 mg arsenic mass was placed in the ampoule to provide As excess pressure. Since Cd and Se diffusion coefficients at 800°C ($D_{Cd} \cong 3 \cdot 10^{16}$ cm$^2 \cdot$s$^{-1}$, $D_{Se} \cong 7 \cdot 10^{17}$ cm$^2 \cdot$s$^{-1}$) are substantially lower than that for arsenic vacancies ($D_{VAs} \cong 1 \cdot 10^{15}$ cm$^2 \cdot$s$^{-1}$), these dopants diffuse in vacuum into the crystals with arsenic deficiency, or into the crystals with arsenic excess under arsenic vapor excess pressure. The reference annealing of the samples without sputtered dopant layers was also carried out in vacuum and under arsenic vapor excess pressure in the same conditions. The emission characteristics of the surface layers for doped crystals were determined in the linear carrier recombination region using the layer-by-layer photostimulated anodic oxidation. The carrier concentration at $n < 10^{16}$ cm$^{-3}$ was determined from the half-width of 300 K band-to-band PL [8], and the conductivity type, by measuring of surface thermo-e.m.f. sign [9].

After annealing of crystals without sputtered dopant layers, a change in $h\nu_m = 0.65$ eV quantum yield at the surface layer was observed, the intensity ratio between the $h\nu_m = 0.63$ eV to $h\nu_m = 0.68$ eV bands remaining unchanged. Annealing in vacuum resulted in decreased $I_{0.65}$ for surface layers (up to 20 μm in depth) (Fig. 2, curve 2). Also, in this case, the surface layers showed p-type conductivity. Such $I_{0.65}$ decrease can be explained by decreasing EL2 center concentration ($[N_{EL2}]$) due to arsenic evaporation from the surface region of samples and formation of $V_{As}$ [2]. At the same time, the reaction of anti-structural defect "decay" according to Eq.(l) is defined by diffusion of arsenic vacancies into the crystal volume. Annealing in arsenic atmosphere caused an increase of $I_{0.65}$ at a depth up to 12 μm from the sample surfaces (Fig. 2, curve 1), the n-type of surface conductivity being retained. This increase is due to formation of EL2 defects as a result of As diffusion [2].

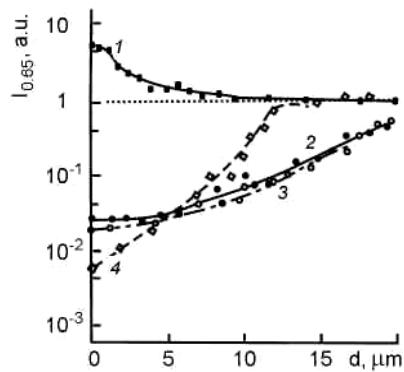

Fig. 2. Emission intensity through EL2 centers after thermal annealing as a function of the distance to crystal surface: *1, 2* — without dopant; *3, 4* — diffusion of Cd atoms; *2, 3* — in vacuum; *1, 4* — in of As vapor atmosphere.

After cadmium diffusion, the crystal surface was characterized by p-type conductivity. In comparison with annealing dopant-free samples, the crystal doping with cadmium in vacuum had no significant impact on quantum efficiency of radiative transitions through EL2 centers (Fig. 2, curve 3). Under excess arsenic pressure, cadmium diffusion resulted in a significant (by 2-3 orders at the crystal surface) decrease of $I_{0.65}$ as compared to typical $I_{0.65}$ values for the case of dopant-free annealing (Fig. 2, curve 4). The nature of quantum yield changes for radiative transitions through EL2 defects in the case of cadmium doping seems to be same that in the case of copper doping [3]. Namely, the reason for $h\nu_m = 0.65$ eV intensity decrease is the lowered concentration of separated EL2 defect resulting from EL2 – Cd complex formation. The fact that the effect is observed only at diffusion in arsenic atmosphere may be explained as follows. As cadmium atoms in GaAs mainly occupy Ga vacancies [6], they can enter the complexes as EL2 – $Cd_{Ga}$ (cadmium diffusion occurs according to dissociative mechanism and the formation probability of EL2 – $Cd_j$- complexes with interstitial atoms is low).

As is noted, the diffusion coefficients for Cd are much lower than for As vacancies. Therefore, cadmium can be believed to diffuse into crystal area where the $V_{As}$ concentration is reduced according to the expression [6]

$$[V_{As}] \cdot [V_{Ga}] = k \cdot p_{As}^{1/2} \qquad (2)$$

(where $p_{As}$ is the As vapor pressure). At the same time, in this area, $V_{Ga}$ concentration (gallium deficiency) increases. Within the frame of our experiment, the impurity diffusion area did not exceed in extent that of the emission quantum yield through EL2 defects at the doping-free annealing. Increased $V_{Ga}$ number near $As_{As}$ defects favors the EL2 – $Cd_{Ga}$ complex formation.

After introduction of Se atoms, the crystal surface showed n-type conductivity. Diffusion in vacuum resulted in a greater decrease of $I_{0.65}$ than in samples annealed without dopant (Fig. 3, curve 3). At the same time, Se diffusion in As atmosphere caused $I_{0.65}$ increase as compared to its values after dopant-free annealing (Fig. 3, curve 4). In conditions of linear carrier recombination, when the carrier lifetime $\tau_n$ in Se-doped region corresponded to that in SIU GaAs, $I_{0.65} \sim N_{EL2}$ and $I_{0.65}$ change is defined by increase or decrease of $N_{EL2}$ [1].

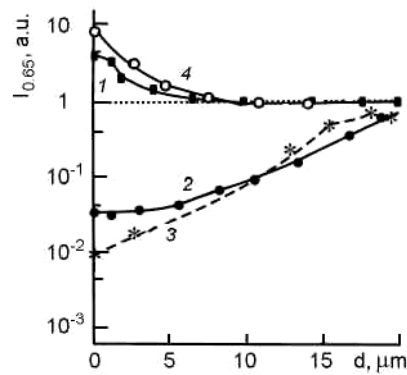

Fig. 3. Emission intensity through EL2 centers after thermal annealing as a function of the distance to crystal surface: *1, 2* — without dopant; *3, 4* — diffusion of Se atoms; *2, 3* — in vacuum; *1, 4* — in As vapor atmosphere.

The increased intensity of emission associ-ated with transitions through EL2 centers after Se diffusion in arsenic atmosphere is caused most likely by partial displacement of As atoms by the impurity introduced in arsenic sublattice [10]. At the same time, the displaced atoms can occupy gallium vacancies, the concentration thereof at crystal surface being increasing during heat treatment in arsenic atmosphere according to the law of active mass [6]. The $I_{0.65}$ decrease resulting from selenium diffusion at vacuum heat treatment can be connected with decreasing efficiency of emission through EL2 defects due to EL2-Se complex formation, as in the cases with cadmium and copper. Thus, diffusion of Se atoms occurs into areas where the concentration of As vacancies $V_{As}$ generated due to arsenic evaporation from the surface is higher than that in the crystal volume. At the same time, the number of As vacancies near $As_{Ga}$ anti-structural defects increases. The latter is possible if the probability of reaction between $As_{Ga}$ and $V_{As}$ in accordance to mechanism (1) is less than one, i.e. there is a potential barrier for recombination thereof. That barrier (exceeding 5 meV at GaAs crystallization temperature) has been evidenced in [11, 12]. As a result of selenium diffusion along vacancies in As sublattice [6], electrically inactive EL2-$Se_{As}$ complexes are formed and the concentration of separated EL2 centers reduces.

To conclude, copper is not the only impurity causing change in quantum efficiency of emission through EL2 defects in GaAs crystals. Introduction of cadmium as an acceptor dopant results in decreased intensity of the corresponding radiation band for crystals with gallium deficiency, while in the case of GaAs crystals with arsenic deficiency the introduction of cadmium does not change intensity of that band. This effect is caused by reduction of separated EL2 centers concentration due to EL2-$Cd_{Ga}$ complex formation. Introduction of selenium donor dopant increases the quantum yield for emission caused by transitions through EL2 defects in crystals with gallium deficiency and decreases it in crystals with arsenic deficiency. The first effect is caused most likely by increasing EL2 center concentration due to displacement of As atoms from arsenic sublattice by the impurity while the second one, by EL2-$Se_{As}$ complex formation.